\documentclass[sigconf, nonacm]{acmart}

\AtBeginDocument{%
  }

\usepackage[utf8]{inputenc}

\usepackage{multirow}
\usepackage{booktabs}
\usepackage{enumitem}
\usepackage{bbold} 
\usepackage{pifont}
\newcommand{\cmark}{\ding{51}}%
\newcommand{\xmark}{\ding{55}}%
\usepackage{ulem}

\usepackage[bb=boondox,bbscaled=.95,cal=boondoxo]{mathalfa}

\begin{document}

\title{Machines in the Crowd? Measuring the Footprint of Machine-Generated Text on Reddit}

\author{Lucio {La Cava}}
\email{lucio.lacava@dimes.unical.it}
\orcid{https://orcid.org/0000-0003-3324-0580}
\affiliation{%
  \institution{DIMES Dept., University of Calabria}
  \country{Italy}
}

\author{Luca Maria Aiello}
\email{luai@itu.dk}
\orcid{https://orcid.org/0000-0002-0654-2527}
\affiliation{%
  \institution{IT University of Copenhagen\\Pioneer Centre for AI}
  \country{Denmark}
}

\author{Andrea Tagarelli}
\email{tagarelli@dimes.unical.it}
\orcid{https://orcid.org/0000-0002-8142-503X}
\affiliation{%
  \institution{DIMES Dept., University of Calabria}
  \country{Italy}
}

\renewcommand{\shortauthors}{La Cava et al.}

\begin{abstract}

Generative Artificial Intelligence is reshaping online communication by enabling large-scale production of Machine-Generated Text (MGT) at low cost. While its presence is rapidly growing across the Web, little is known about how MGT integrates into social media environments. In this paper, we present the first large-scale characterization of MGT on Reddit. Using a state-of-the-art statistical method for detection of MGT, we analyze over two years of activity (2022–2024) across 51 subreddits representative of Reddit's main community types such as information seeking, social support, and discussion. We study the concentration of MGT across communities and over time, and compared MGT to human-authored text in terms of social signals it expresses and engagement it receives. Our very conservative estimate of MGT prevalence indicates that synthetic text is marginally present on Reddit, but it can reach peaks of up to 9\% in some communities in some months. MGT is unevenly distributed across communities, more prevalent in subreddits focused on technical knowledge and social support, and often concentrated in the activity of a small fraction of users. MGT also conveys distinct social signals of warmth and status giving typical of language of AI assistants. Despite these stylistic differences, MGT achieves engagement levels comparable than human-authored content and in a few cases even higher, suggesting that AI-generated text is becoming an organic component of online social discourse. This work offers the first perspective on the MGT footprint on Reddit, paving the way for new investigations involving platform governance, detection strategies, and community dynamics.
\end{abstract}

\maketitle

\section{Introduction}
\label{sec:intro}

Generative Artificial Intelligence (GenAI) is transforming the Web and reshaping everyday practices related to search~\cite{white2024advancing}, news consumption~\cite{pavlik2023collaborating}, entertainment~\cite{anantrasirichai2022artificial}, and online social interactions~\cite{bail2024can}.
In particular, the dynamics of the participatory Web are being revolutionized by the commercial release of Large Language Models (LLMs) that dramatically lowered the barriers to producing Machine-Generated Text (MGT).
The broad diffusion of these new tools is paving the way for large volumes of crowd-generated synthetic content to be uploaded to social media and other participatory online platforms.

This socio-technical shift carries profound societal implications.
The Web—and especially social media—has become a central arena for public discourse and information sharing, shaping the spread of ideas and influencing democratic participation.
The diffusion of MGT may significantly affect these collective processes~\cite{bengio2024managing} in ways that scholars are still actively debating.
Optimistic perspectives emphasize the potential of GenAI to enhance contributors’ creativity~\cite{doshi2024generative}, increase engagement, and promote inclusivity in online discussions~\cite{tessler2024ai}.
In contrast, critics raise concerns about misinformation, declining authenticity in user interactions~\cite{tessler2024ai,floridi2020gpt},deteriorating content quality~\cite{floridi2020gpt,moller2025impact}, and emergent biases~\cite{ashery2025emergent}.
When produced with strategic or malicious intent, MGT can be deliberately crafted to be persuasive~\cite{breum2024persuasive} and highly engaging~\cite{schroeder2025malicious}, while embedding inaccurate, biased, or harmful information~\cite{floridi2020gpt}.
This could exacerbate systemic risks identified by the European Digital Services Act, including discrimination, declining mental well-being, and the erosion of civic and electoral processes~\cite{turillazzi2023digital}.
Developing a quantitative understanding of AI’s impact on people’s online experiences is therefore both essential and urgent to inform effective policies and design safeguards against potential harms.

Although distinguishing Machine-Generated Text (\textbf{MGT}) from Human-Generated Text (\textbf{HGT}) text is inherently challenging, promising tools have been developed with high accuracy and inference speed~\cite{BaoZTY024}.
These tools have enabled initial estimates of the proliferation of synthetic content online~\cite{sun2024we} and of the growing presence of AI agents posing as human users~\cite{schroeder2025malicious}.
However, beyond these initial estimates, the use of MGT in real-world online social environments remains mostly unexplored.
To shed light on how online social dynamics might change as result of MGT, it is not only important to estimate its incidence, but also to analyze the nature of the content produced, the context in which it is published, and the response that the public has to it.

To help address this gap, we present the first large-scale characterization of MGT on Reddit, one of the world’s leading social media platforms. 
With its millions of active users and countless topic-based discussions, Reddit stands as a compelling case in point for exploring whether and to what extent MGT emerges and spreads  with human discourse online.
Focusing on 51 popular subreddits representative of Reddit’s main functional community types, we analyze comments and submissions from 2022 to 2024, a period that includes   milestones in the release of GenAI tools. We used a state-of-the-art method to detect messages with a high likelihood of being MGT and address four main research questions:  

\begin{enumerate}[leftmargin=2.5em]
\item[\textbf{RQ1}] --- 
How is MGT adopted across subreddit communities and over time, and how does this reflect the distinct conversational norms of different community categories?
\item[\textbf{RQ2}] --- 
What temporal patterns characterize MGT prevalence, and how might these relate to exogenous events?
\item[\textbf{RQ3}] --- 
How does the distribution of MGT users evolve over time, and are there any significant trends in its adoption?
\item[\textbf{RQ4}] --- 
Do MGT and HGT differ in the type or intensity of social signals and engagement patterns they convey across subreddit categories?
\end{enumerate}

\section{Methodology}
\label{sec:methods}

We present the data (Section~\ref{sec:methods:data}), the technique for estimating the likelihood of text being MGT (Section~\ref{sec:methods:detection}), and the methods for assessing the nature of the content (Section~\ref{sec:methods:content}) and engagement (Section~\ref{sec:methods:engagement}) in MGT and HGT.

\subsection{Data}
\label{sec:methods:data}
To characterize the emergence and diffusion of AI-generated texts on Reddit, we start by considering the top 1,000 subreddits by number of subscribers.\footnote {\url{https://www.reddit.com/best/communities/1/}} 
The choice of analyzing large and active communities is motivated by (i) the broad visibility and potential impact that the content posted in those communities have on the public, and (ii) the substantial volume of data providing sufficient support for a reliable estimation of the prevalence of MGT.

We manually parsed the list of subreddits in decreasing order of popularity and selected a representative subset of 51 subreddits that we mapped into five main functional subreddits categories that have been informed by a taxonomy form prior work~\cite{WeldZA22}:

\begin{itemize}[leftmargin=*]
\raggedright
\item \textbf{Information Seeking}, including, e.g., r/worldnews, r/askscience, r/explainlikeimfive, r/health; 
\item \textbf{Social Support}, including, e.g., r/GetMotivated, r/mentalhealth, r/AITAH, r/LifeProTips;
\item \textbf{Discussion}, including, e.g., r/changemyview, r/unpopularopinion, r/SeriousConversation, r/politics;
\item \textbf{Identity}, including, e.g., r/teenagers, r/asktransgender, r/AskWomen, r/BlackPeopleTwitter;
\item \textbf{Chit Chat}, including, e.g., r/funny, r/entertainment, r/books, r/Showerthoughts.
\end{itemize}

We collected the complete data on submissions and comments posted in these subreddits from January 2022 to December 2024 (see Appendix~\ref{app:taxonomy} for the complete list of subreddits) using data dumps obtained from the \textit{PushShift API}~\cite{pusshift}. This covers the period during which LLMs and GenAI have become widely accessible to the public, allowing us to study their adoption and diffusion in online discourse through Reddit. After filtering out empty posts (e.g., containing only images or URLs), we collected \textbf{38,074,021 comments} and \textbf{4,073,586 submissions}. 
A detailed overview of the number of analyzed comments for each month is reported in Figure~\ref{fig:appendix-temporal-distrib}, Appendix~\ref{app:temporal-distribution-comments}.

\subsection{Detecting Machine-Generated Text}
\label{sec:methods:detection}

\vspace{1mm}
\noindent 
\textbf{Problem setting.\ }
We are given a collection of text messages (e.g., Reddit comments  or submissions)  $\mathcal{X} = \{x_1, ..., x_n\}$ where each message $x_i$ can be authored by either a human (i.e., HGT) or any machine-generator tool like large language models (i.e., MGT), but the type of author is unknown.  
We frame the detection of MGT content in online discussions  
as a \textit{binary text classification task}: given a classifier model $f$, the goal is to assign a label $y_i \in \{0,1\}$ to each $x_i \in \mathcal{X}$ such that:
\begin{equation}
y_i =
\begin{cases}
1, & \text{if } P(MGT \mid x_i) \geq \tau \\
0, & \text{otherwise},
\end{cases}
\label{eq:task}
\end{equation}
where $P(MGT \mid x_i)$ denotes the probability, estimated by $f$, that $x_i$ is an MGT, and $\tau \in [0,1]$ is a decision threshold, ideally selected to reflect the intended trade-off between false positives and false negatives.

\vspace{1mm}
\noindent 
\textbf{Approach.\ }
To detect signals of generative AI usage within Reddit communities, we considered two major classes of MGT detection techniques:
\begin{itemize}[leftmargin=*]
    \item \textit{Metric-based} approaches~\cite{solaiman2019release, gehrmann-etal-2019-gltr, pmlr-v202-mitchell23a, su-etal-2023-detectllm, BaoZTY024} rely on statistical properties of texts such as token distributions, entropy measures, perplexity;
    \item \textit{Model-based} approaches~\cite{solaiman2019release, guo2023detector, guo2024detective, whosai} leverage deep-learning classifiers trained for distinguishing HGT from MGT.
\end{itemize}

Although both approaches have been widely adopted in the literature and demonstrated strong performance~\cite{He2024mgtbench}, we opted for a metric-based solution because of two main practical constraints. 
First, model-based approaches are typically trained on datasets from domains like news articles or essays, leading to degraded performance on conversational data such as Reddit comments. Moreover, fine-tuning such approaches on Reddit data would be costly in practice, as it would require a large training set of labeled Reddit-specific data---currently unavailable, to the best of our knowledge. 
Second, model-based approaches require relatively slow (neural) inference for each input text, which becomes computationally prohibitive at a scale of millions of comments and submissions.

Based on these considerations, we selected the zero-shot metric-based \textit{Fast-DetectGPT}~\cite{BaoZTY024}, as our primary detection tool. Fast-Detect\-GPT strikes a good balance between detection capabilities and computational efficiency, ranking among the most effective detectors in recent benchmarks~\cite{otb} while remaining significantly faster than model-based detectors and commercial alternatives (e.g., \textit{GPTZero}).
\textit{Fast-DetectGPT} operates by measuring the divergence between the log-probabilities (extracted through a small, efficient language model) of a text and the distribution of perturbed variants of the same text, based on the hypothesis that a significant shift would follow typical statistical patterns of machine-generated texts~\cite{BaoZTY024}.

\subsection{Content Analysis}
\label{sec:methods:content}

\vspace{1.5mm}
\noindent \textbf{Syntactic and Readability Characteristics.\ }
We measure  main features of HGT and MGT messages, focusing on style and linguistic  aspects of the text such as 
(i) number of words and sentences,  
(ii) compression ratio  (i.e., original size divided by gzip compressed size), and 
(iii) Flesch Reading Ease score indicating the degree of difficulty in comprehending a text passage as a function of the number of words, syllables, and sentences it contains; the score ranges within -$\infty$ (low readability) and $121.22$ (high readability).\footnote{\url{https://pypi.org/project/textstat/}} 

\vspace{1.5mm}
\noindent \textbf{Social Dimensions.\ }
We compare MGT and HGT in terms of the strength of the \textit{social intent} they convey. Social theorists have identified universal hallmarks of communication that capture fundamental dimensions of intent emerging in social interactions, such as offering support, exchanging knowledge, or expressing conflict~\cite{Deri2018coloring}. These dimensions generalize across topics and provide a \textit{functional} characterization of the text, which we expect to vary across community types and between MGT and HGT. Prior work introduced a method for automatically identifying these social dimensions in conversational text~\cite{Choi20tensocial}.\footnote{Available as a Python package at \url{https://github.com/lajello/tendimensions}}  This tool consists of a set of classifiers (one per dimension) trained and previously validated on Reddit comments (average area under the ROC curve of 0.84 across dimensions), making them directly applicable to our study.

We use these classifiers according prior work's best practices~\cite{monti2022language}. Given a message $x$ and a social dimension $d$, the classifier outputs a score $\phi_d(sen) \in [0,1]$ for each sentence $sen \in x$, representing the likelihood that $sen$ conveys dimension $d$. In our setting, we associate each message $x$ with the maximum score across its sentences, i.e., $\phi_d(x) = \max_{sen \in x} \phi_d(sen)$. Using the maximum rather than the average avoids dilution effects and aligns with the theoretical interpretation of social dimensions being effectively conveyed even if expressed when expressed only briefly in a message~\cite{Deri2018coloring}.

To identify messages that convey a given dimension with high probability, we discretize the classifier outputs into binary indicators. For each message $x$ and dimension $d$, we consider $x$ to convey $d$ if $\phi_d(x)$ exceeds a dimension-specific threshold $\theta_d$. Favoring precision over recall, we set $\theta_d$ to the $85^{\text{th}}$ percentile of observed scores for each classifier. Because longer messages have a higher chance of containing at least one sentence above the threshold, we mitigate the length bias with a discounting factor. Specifically, we standardize the length of each message $x$, denoted $\ell_d(x)$, with respect to the distribution of message lengths for dimension $d$: $\widehat{\ell}_d(x) = \dfrac{\ell_d(x)-\mu_d}{\sigma_d}$, where $\mu_d$ and $\sigma_d$ represent the mean and standard deviation of lengths across all messages conveying $d$. Finally, to give greater weight to shorter messages that strongly express a given dimension, while down-weighting cases where the dimension appears only due to excessive length, we define the \textit{social dimension score} of $x$ with respect to $d$, denoted $\varphi_d(x)$, as follows:
\begin{equation}
    \varphi_d(x) = 
    \begin{cases}
        \dfrac{1}{1+\widehat{\ell}_d(m)} & \text{if } \phi_d(x) \geq \theta_d  \; \wedge \; \widehat{\ell}_d(m)\geq 0\\
        
        2-\dfrac{1}{1-\widehat{\ell}_d(m)} & \text{if } \phi_d(x) \geq \theta_d \; \wedge \; \widehat{\ell}_d(m)<0\\

        0 & \text{if } \phi_d(x) < \theta_d. 
    \end{cases}
\label{eq:finaldim}
\end{equation}

The score ranges in $[0,2]$, and is equal to 1 when the text length is as the average for that dimension, approaches 0 as the text becomes substantially longer, and 2 as it becomes shorter. 

For our study, and given the categories identified in Table~\ref{tab:data-stats}, we considered the dimensions that are most relevant to the Reddit discourse: \textit{knowledge}, \textit{status}, \textit{support}, \textit{fun}, \textit{conflict}, and \textit{similarity} (cf. Appendix~\ref{app:socialdimensions} for details on each dimension).

\begin{table}[t!]
    \centering
    \caption{Dataset statistics across subreddit categories after filtering. Count indicates the number of subreddits within each category.}
    \label{tab:data-stats}
    \scalebox{0.9}{
    \begin{tabular}{lcrr}
    \toprule
    \textbf{Category} & \textbf{Count} & \textbf{\# Submissions} & \textbf{\# Comments} \\
    \midrule
    Information Seeking & 13 & 246,106 & 1,635,872 \\
    Social Support & 18 & 1,716,198 & 4,541,287 \\
    Discussion & 5 & 42,371 & 1,453,938 \\
    Identity & 10 & 99,855 & 615,020 \\
    ChitChat & 5 & 26,029 & 785,886 \\
    \midrule
    Total & 51 & 2,130,559 & 9,032,003 \\
    \bottomrule
    \end{tabular}
    } 
\end{table}

\subsection{Engagement Analysis}
\label{sec:methods:engagement}

The public engagement generated by messages in online communities is reflected in the number of \textit{upvotes} and \textit{downvotes} they receive.
For any message $x_i$, we use the difference between upvotes and downvotes as a direct measure of its positive engagement, which we refer to as the \textit{engagement score} ($e_i$).

Our goal is to compare the engagement levels of MGT with those of HGT. 
  To ensure a sufficiently fine-grained resolution, we organized the data on timespans a \textit{monthly} basis and at the \textit{subreddit level}. 
 Hereinafter, we denote with $I=\{(s,t)\}$ the set of (subreddit, timespan) instance pairs under examination. 
We assume that messages posted within a subreddit during a given timespan may include some fraction of MGT, although we expect this fraction to be substantially smaller than that of HGT.
To compare engagement within a fixed context, we restrict our analysis to each subreddit–timespan pair $(s,t) \in I$ individually.
For each pair, we first identify the MGT messages, and then draw $N = 1{,}000$ bootstrap samples (with replacement) from the pool of HGT. 
Each of the $N$ samples has the same size as the set of MGT messages detected for that subreddit and timespan.
We denote by $B_{(s,t)}^{(M)}$ the set of MGT messages detected for $(s,t)$, and by $B_{(s,t)}^{(j)}$ the $j$-th bootstrap sample of HGT messages, with $j = 1 \dots N$.

Given a bootstrap sample of HGT messages ($B_{(s,t)}^{(j)}$) and the corresponding set of MGT messages ($B_{(s,t)}^{(M)}$), we compared their engagement score distributions to assess the significance and magnitude of their difference. To do so, we employed the nonparametric \textit{Mann–Whitney U test} for comparing numerical distributions. Its null hypothesis is that values from the two groups are drawn from the same distribution, and the test evaluates this by comparing the ranks of the observations. If the resulting p-value falls below the significance threshold of $0.05$, we reject the null hypothesis, indicating a statistically significant difference in engagement between MGT and HGT.

For each subreddit–timespan pair $(s,t)$, we conducted $N$ two-sided Mann–Whitney tests (one per bootstrap sample). Focusing only on the samples $B_{(s,t)}^{(j*)}$ for which the null hypothesis is rejected, we estimated the effect size relative to $B_{(s,t)}^{(M)}$ using \textit{Cliff’s delta} ($\delta$), defined as:
\begin{equation}
    \delta_{B_{(s,t)}^{(j*)}} = \frac{1}{k^2} \sum_{x_i \in B_{(s,t)}^{(j*)}} \big (\mathbb{1}[e_i^{\textrm{MGT}} > e_i^{\textrm{HGT}} ] - \mathbb{1}[e_i^{\textrm{HGT}} > e_i^{\textrm{MGT}} ] \big )
\end{equation}
\noindent 
where $k$ is the number of MGT messages detected for $(s,t)$; $e_i^{\textrm{MGT}}$ and $e_i^{\textrm{HGT}}$ denote the engagement scores of the $i$-th MGT and human-written message, respectively; and $\mathbb{1}$ is the indicator function. The term inside the sum reflects the difference between the number of pairwise comparisons in which MGT comments receive higher, versus lower, engagement scores than human-written comments. Cliff’s $\delta$ ranges in $[-1,1]$, where $\delta > 0$ indicates that MGT comments receive higher engagement, $\delta < 0$ indicates the opposite, and $\delta$ close to $0$ suggests negligible differences in engagement.

\section{Results}
\label{sec:results}

\noindent 
\textbf{Experimental Setting.\ }
To ensure that our findings are robust and that the detected signals are not driven by noise, we adopt a conservative detection policy. Specifically, we set the detection threshold $\tau$ in Equation~\ref{eq:task} to 0.99, and restrict our analysis to texts (both submissions and comments) containing at least 250 tokens; it should be noted that the choice of this token length threshold is supported by evidence  provided in~\cite{BaoZTY024}.

Table~\ref{tab:data-stats} provides an overview of the resulting dataset under these criteria. Although this choice narrows our focus to \textbf{9,032,003 comments} (with an average length of $387.39 \pm 180.16$ tokens) and \textbf{2,130,559 submissions} (with an average length of $542.79 \pm 331.87$ tokens), it is motivated by two key considerations. First, using a high threshold $\tau$ in the reference detection method substantially reduces the risk of false positives, ensuring that texts labeled as machine-generated represent highly confident detections. Second, prior work has shown that detection reliability improves with text length~\cite{ChakrabortyBZAM24}. 
Taken together, these constraints provide conservative yet trustworthy evidence of the presence of MGT on Reddit.

We investigated the prevalence of MGT in \textit{both} submissions and comments on Reddit. 
Nonetheless, due to space limitations, we focus our presentation on comments for three main reasons: they provide clear evidence of interaction flow, they are more likely to reflect genuine communication, and we empirically observed that comments exhibit lower sparsity over time, making them better suited for fine-grained temporal analysis.

 Therefore, in the following, we  address the previously stated RQs 
 across subreddit comments, i.e.,  \textit{where} and \textit{when} MGT is used, \textit{who} adopts it, and \textit{what} types of content are most likely to be identified as MGT.  
 \emph{We replicate the same experimental analysis on the submissions and report it   in Appendix~\ref{app:submissions-results}}.

\begin{table}[t]
\centering
\caption{Peak prevalence of MGT across subreddit categories. For each category, we report the fraction of MGT comments over the total detected, the maximum observed share of MGT, the subreddit in which it occurred, and the corresponding date (yy/mm).}
\label{tab:mgt-peaks}
\scalebox{0.9}{
\begin{tabular}{lrrlc}
\toprule
\textbf{Community} & \textbf{MGT \%} & \textbf{Peak \%} & \textbf{Top Subreddit} & \textbf{Date} \\
\midrule
Inf. Seek.   & 29.83       & 6.33 & \texttt{r/askscience}          & 23/07 \\
Soc. Supp.   & 26.73       & 7.69 & \texttt{r/malefashionadvice}   & 23/07 \\
Discussion   & 20.83       & 1.28 & \texttt{r/politics}            & 23/09 \\
Identity     & 17.81       & 8.46 & \texttt{r/teenagers}           & 23/02 \\
ChitChat     & 4.80       & 3.13 & \texttt{r/funny}               & 23/05 \\
\bottomrule
\end{tabular}
}
\end{table}

\begin{figure}[ht!]
    \centering
    \includegraphics[width=0.97\linewidth]
    {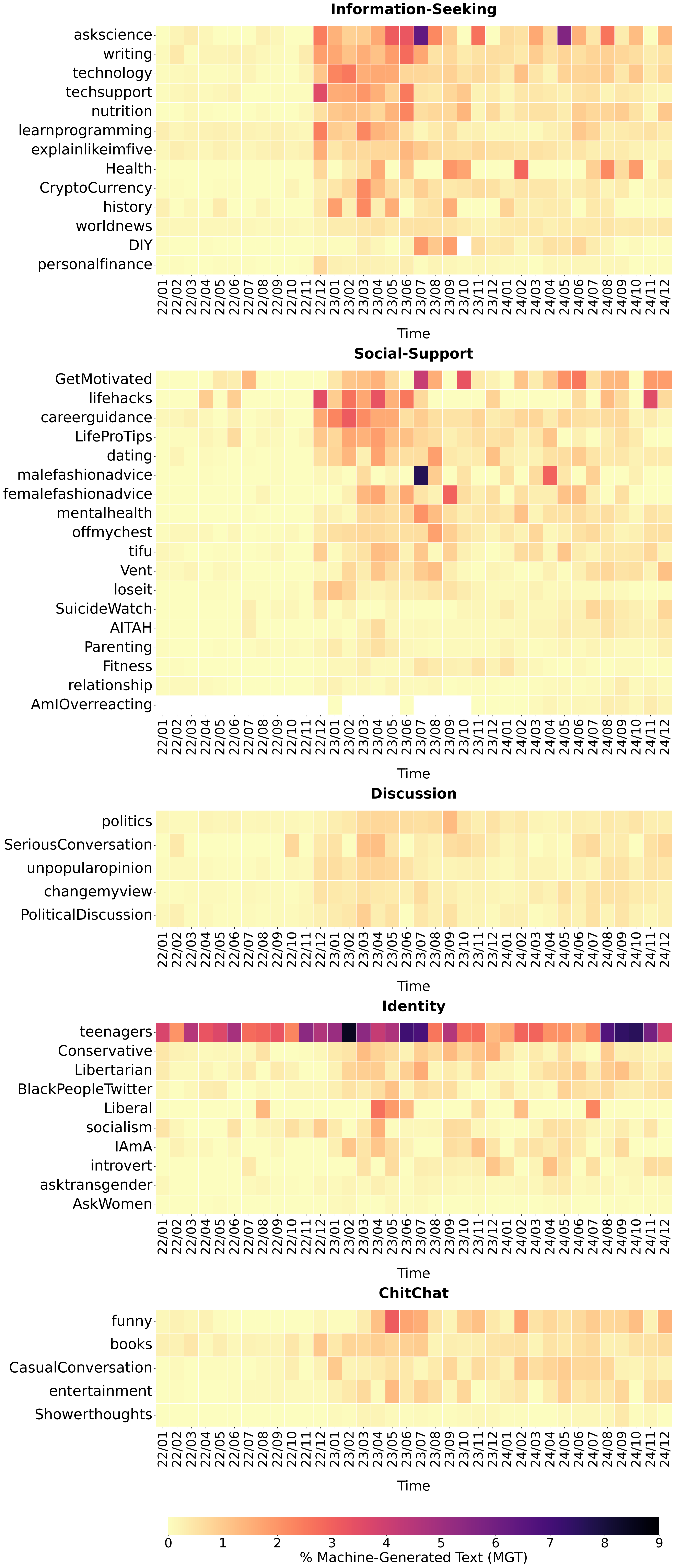}
    \caption{MGT comment usage across subreddits. Rows are sorted by the sum of monthly MGT adoption. Each cell shows the share of MGT posts for a subreddit-month pair. Darker shades indicate higher usage. Empty cells mean no comments matching our filtering criteria.}
    \label{fig:rq1}
\end{figure}

\subsection{Community Distribution}
\label{sec:results:where}
The adoption of MGT on Reddit varies considerably across subreddits, reflecting differences in conversational norms determined by the different community categories (cf. Sect. \ref{sec:methods:data}). We summarize results in Table~\ref{tab:mgt-peaks} and Figure~\ref{fig:appendix-temporal-distrib}, and elaborate on them next. 

Information-Seeking communities exhibit among the highest and most consistent adoptions of MGT, with subreddits like \texttt{askscience} showing clear spikes in MGT contributions, peaking at 6.33\% in July 2023. This is supported by the observation that such communities are structured around question-answering dynamics, where GenAI is expected to produce helpful responses with minimal effort, thus lowering the interaction barrier.

Social-Support communities show moderate yet unevenly distributed adoption across subreddits, with some more consistent usage in spaces like \texttt{GetMotivated} or \texttt{lifehacks}, and rare yet significant peaks in others like \texttt{malefashionadvice} (up to 7.69\% of MGT in July 2023). Interestingly, subreddits addressing more sensitive topics (e.g., \texttt{SuicideWatch} or \texttt{mentalhealth}) exhibit minimal MGT activity. This hints at more spontaneous interactions, likely driven by community norms.

Notably, discussion-oriented communities consistently exhibit the lowest adoption of MGT, with only sporadic and slight signals of usage in communities like \texttt{politics} and \texttt{SeriousConversation}, with the former peaking at no more than 1.28\% of MGT in September 2023. We ascribe this phenomenon to the nature of these spaces, where users typically want to express their own views and arguments, rather than resorting to GenAI to generate them.

The identity-focused communities display the most noteworthy outlier pattern: the \texttt{teenagers} subreddit exhibits consistently high adoption of MGT (up to 8.46\%) across all analyzed periods, rendering them as early-adopters of GenAI tools. This phenomenon is likely supported by (i) higher curiosity and willingness to experiment with emerging technologies, (ii) faster peer-to-peer diffusion via social networks, and (iii) lower barriers to adoption due to higher familiarity with digital tools. Moreover, other identity-based subreddits, like politically driven ones, exhibit observable adoption of MGT.

Finally, for ChitChat communities, we observe lower adoption of MGT compared to other spaces, with some slight peaks at 3.13\% for \texttt{funny} in May 2023. We ascribe this more contained prevalence to the low-effort and more spontaneous nature of the corresponding subreddits.

\subsection{Temporal Patterns}
\label{sec:results:when}

Figure~\ref{fig:appendix-temporal-distrib} shows that MGT adoption follows clear temporal patterns. We link these fluctuations to exogenous events, particularly major GenAI milestones and the release of new tools during the observation period of our study. 

The first measurable increase in machine-generated text appears in November 2022, coinciding with the release of \textit{ChatGPT}. From this point onward, most subreddits in Figure~\ref{fig:appendix-temporal-distrib} display increasing prevalence of MGT, as GenAI tools transitioned from software packages requiring programming expertise to accessible chatbot interfaces, thus lowering barriers to adoption. A notable exception is the \texttt{teenagers} subreddit, where modest MGT signals are detected even before the widespread diffusion of ChatGPT. Manual inspection suggests that these early MGT traces are of very low quality (e.g., lacking punctuation, containing repetitive patterns), and likely stem from preliminary statistical tools predating widely usable models such as GPT.

The months immediately following ChatGPT's release mark a phase of rapid adoption across nearly all subreddits, producing the strongest concentration of MGT in the entire observation period. This adoption ``wall'' is followed by several peaks that may correspond to later LLM releases. For instance, the higher prevalence observed in Q1–Q2 2023 in communities such as \texttt{askscience} and \texttt{teenagers} coincides with the release of GPT-4 (March 2023). Thereafter, adoption patterns diverge across communities, with some consolidating a persistent prevalence of MGT, while others show a decline in signals. A renewed increase in adoption appears in Q1–Q2 2024, coinciding with major releases such as \textit{Gemini}, \textit{Claude}, \textit{Grok}, and \textit{Llama}.

\begin{figure}[t!]
    \centering
    \includegraphics[width=1\linewidth]{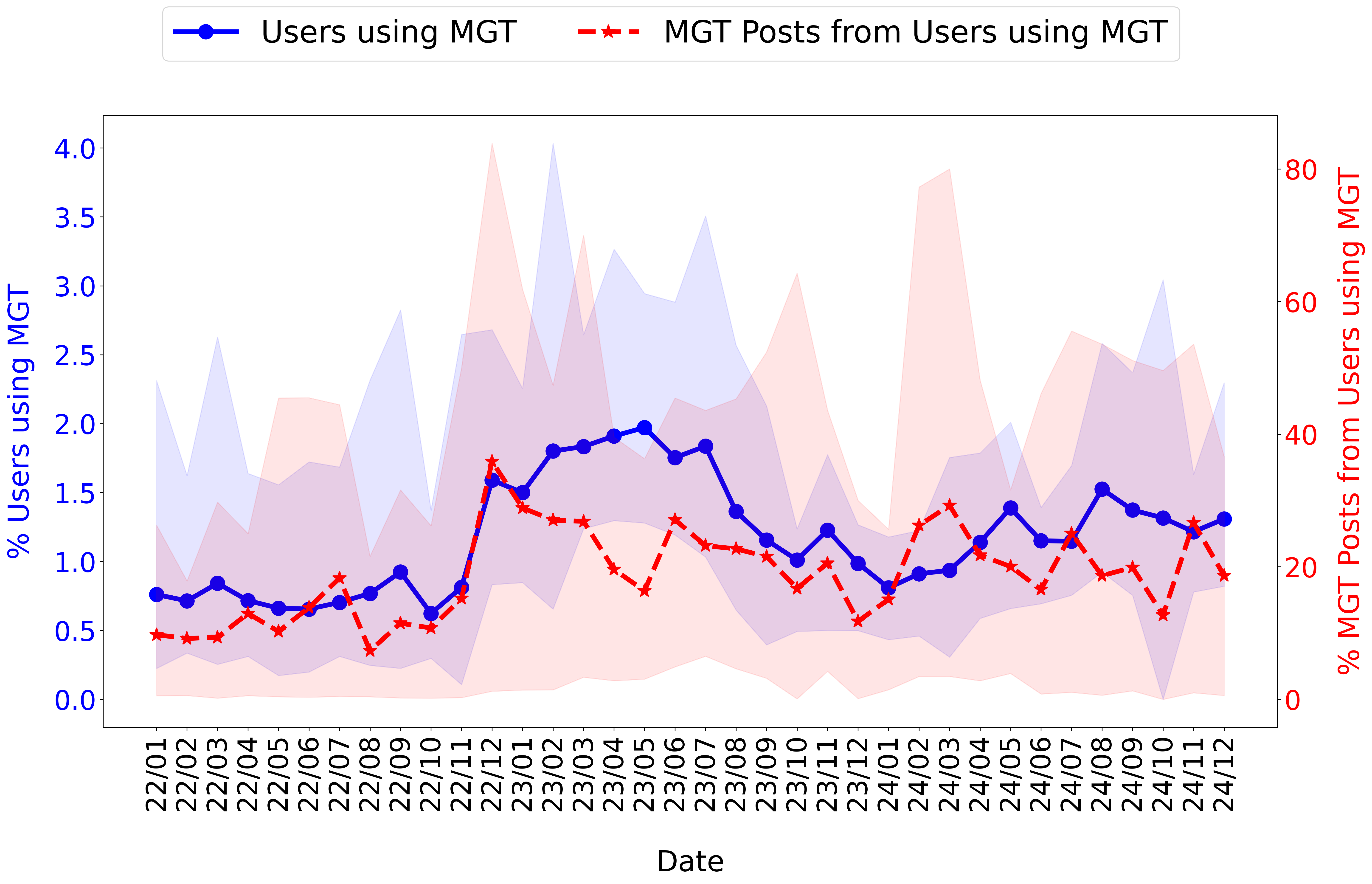}
    \caption{Average  number of users adopting MGT (in blue, left y-axis) and the corresponding percentage of their comments detected to be MGT (in red, right y-axis) across all categories over time. Shaded areas denote minimum and maximum observed values. Users with a single post have been filtered out to mitigate noise due to one-off activities.}

    \label{fig:users_posts_aiusage}
\end{figure}

\subsection{User Distribution}
\label{sec:results:who}

To understand the evolution in adoption of MGT among Reddit users, we analyzed the average fraction of users adopting MGT across the observation period, as well as the percentage of their comments that are detected as being machine-generated. 

Overall, Figure~\ref{fig:users_posts_aiusage} suggests that the adoption of MGT by Reddit users remained relatively low over time, with mean values peaking at 2\% and maximum values reaching 3\% in certain periods. Notably, temporal dynamics reveal distinct adoption phases. In particular, after a slight signal of usage traceable to early adopters, a marked increase is detected in correspondence with the widespread release of major GenAI tools, suggesting a likely relationship between tools' accessibility and adoption. Interestingly, this peak led to the highest rate of adoption across our observation period. After this, adoption stabilized at slightly elevated levels compared to the early adoption period, suggesting the establishment of a small yet persistent cohort of MGT-assisted users.

The fraction of MGT content that these users produce compared to their total comment production is highly heterogeneous, ranging between 10\% and 40\% of posts. Notably, the fraction of MGT posts remains substantial (i.e., around 20\%) even after the initial phase adoption. This suggests that, once incorporating GenAI tools into their workflow, such users tend to rely on them for a significant share of their produced content.

\subsection{Content Features}
\label{sec:results:what}

\vspace{1mm}
\noindent 
\textbf{Text Statistics.\ }
Table~\ref{tab:statistics} summarizes aggregated textual statistics, providing a quantitative comparison of the structural properties of MGT and HGT. Across all subreddit categories, MGT is consistently longer than HGT. Moreover, MGT is generally less readable, with this difference particularly pronounced in Identity-oriented communities. We attribute this to the very low-quality content observed in \texttt{r/teenagers} (e.g., lack of punctuation, highly repetitive patterns), which impacts the aggregated readability scores for the category. Additionally, MGT is more compressible than human comments, reflecting a higher degree of internal redundancy or repetitive phrasing. Overall, MGT comments are typically more verbose than HGT and slightly more difficult to read.

\begin{table}[t!]
\caption{Aggregated values of text statistics from the MGT and HGT comments across subreddit categories. All category–statistics combinations exhibit statistically significant distributional differences ($p< 0.05$) according to the Mann–Whitney U test.}
\label{tab:statistics}
\setlength{\tabcolsep}{1.8pt}
\scalebox{0.8}{
\begin{tabular}{lc|r|r|r|r}
\toprule
\textbf{Category} & \textbf{MGT} 
& \multicolumn{1}{c|}{\textbf{\# Words}} & \multicolumn{1}{c|}{\textbf{\# Sentences}} & \multicolumn{1}{c|}{\textbf{Flesch Read.}} & \multicolumn{1}{c}{\textbf{Compression}} \\
\midrule

\multirow{2}{*}{\textbf{Inf. Seek.}} & {\color{black}\xmark} & 263.94 $\pm$ 167.29 & 15.14 $\pm$ 9.67 & 52.27 $\pm$ 29.06 & 2.94 $\pm$ 0.31 \\
& {\color{black}\cmark} & 334.77 $\pm$ 187.36 & 18.59 $\pm$ 11.95 & 52.46 $\pm$ 58.32 & 3.45 $\pm$ 4.45 \\
\midrule

\multirow{2}{*}{\textbf{Soc. Supp.}} & {\color{black}\xmark} & 264.82 $\pm$ 124.07 & 16.93 $\pm$ 8.71 & 64.43 $\pm$ 21.96 & 2.9 $\pm$ 0.27 \\
& {\color{black}\cmark} & 339.12 $\pm$ 154.21 & 19.75 $\pm$ 10.54 & 63.58 $\pm$ 33.88 & 3.25 $\pm$ 1.23 \\
\midrule

\multirow{2}{*}{\textbf{Discuss.}} & {\color{black}\xmark}  & 253.39 $\pm$ 138.77 & 14.59 $\pm$ 7.58 & 50.63 $\pm$ 25.72 & 2.92 $\pm$ 0.31 \\
& {\color{black}\cmark} & 418.88 $\pm$ 232.35 & 22.32 $\pm$ 14.4 & 48.52 $\pm$ 37.66 & 3.45 $\pm$ 2.4 \\
\midrule

\multirow{2}{*}{\textbf{Identity}} & {\color{black}\xmark}  & 317.39 $\pm$ 226.31 & 16.84 $\pm$ 20.05 & 38.71 $\pm$ 146.23 & 4.73 $\pm$ 11.99 \\
& {\color{black}\cmark} & 459.34 $\pm$ 289.56 & 22.55 $\pm$ 24.32 & -32.96 $\pm$ 280.96 & 8.39 $\pm$ 22.5 \\
\midrule

\multirow{2}{*}{\textbf{ChitChat}} & {\color{black}\xmark}  & 221.42 $\pm$ 131.9 & 13.61 $\pm$ 6.72 & 52.03 $\pm$ 42.12 & 2.97 $\pm$ 2.03 \\
& {\color{black}\cmark} & 348.67 $\pm$ 182.04 & 19.11 $\pm$ 12.74 & 50.31 $\pm$ 94.24 & 3.6 $\pm$ 5.4 \\
\bottomrule
\end{tabular}
} 
\end{table}

\begin{figure*}
    \centering
    \includegraphics[width=1\linewidth]{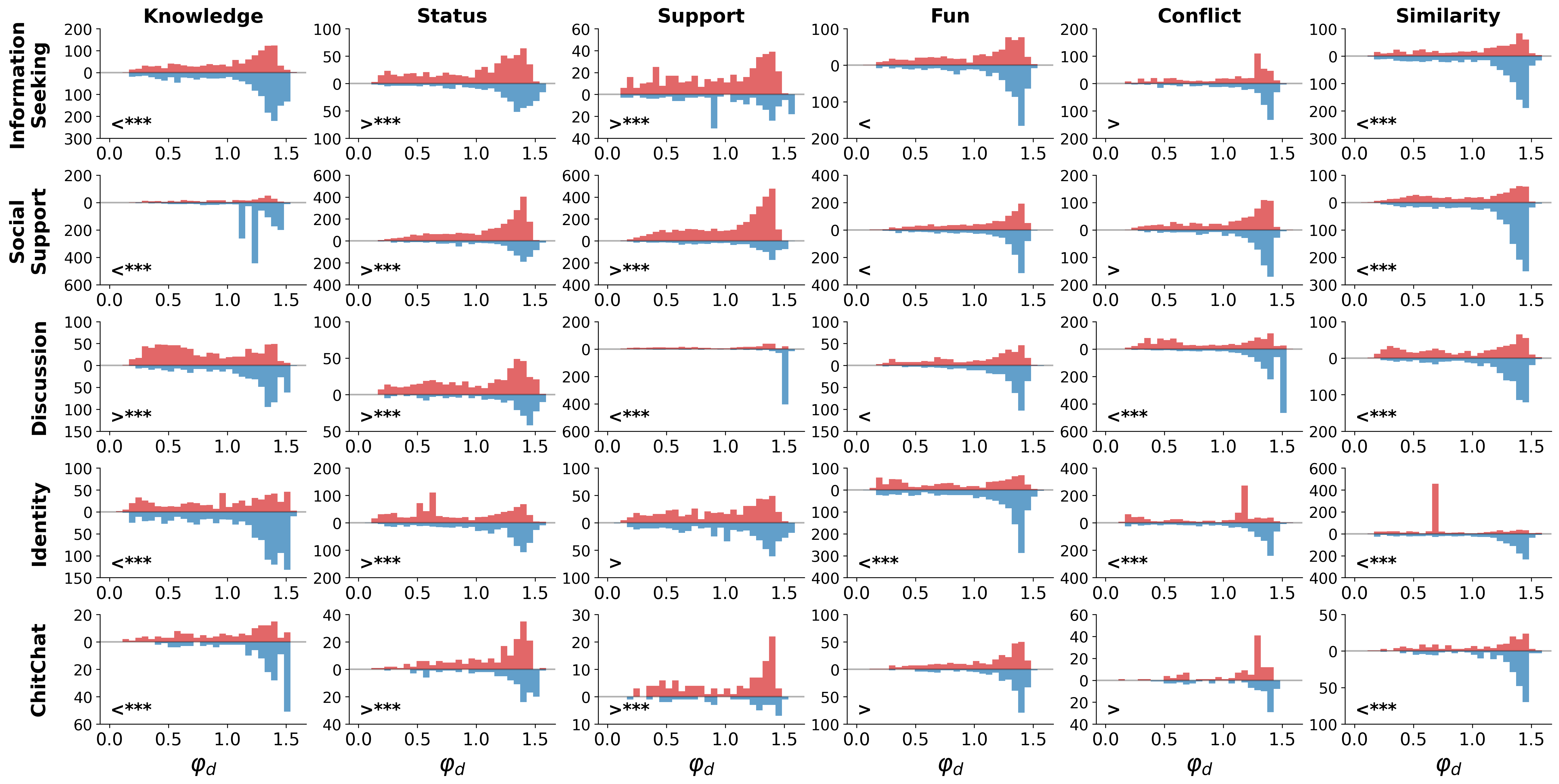}
    \caption{Comparison of social dimension expressions between MGT (in red) and HGT (in blue) comments across different subreddit categories. Each subplot shows the distribution of social dimension scores $\phi_d(x)$ (excluding instances with $\phi_d(x)=0$) for all combinations of subreddit and social dimension. Each subplot is labeled with the corresponding effect size, where $>$ indicates that MGT scores higher than HGT, and $<$ indicates the opposite. Marker $^{***}$   denotes that the difference in social dimension expression scores between MGT and HGT is statistically significant according to the Mann–Whitney U test.}
    \label{fig:tendims}
\end{figure*}

\vspace{1mm}
\noindent 
\textbf{Social Dimensions.\ }
By applying to the social dimension classifier described in Section~\ref{sec:methods}, we performed a qualitative analysis to assess whether and to what extent MGT and HGT differ in the type or strength of social signals they convey across subreddit categories. Figure~\ref{fig:tendims} illustrates the results, grouped for each combination of subreddit category and dimension, and annotated with the corresponding significance and effect size.
Next, we will discuss the statistically significant differences according to a Mann–Whitney U test, and provide interpretations on a per-category basis. 

\vspace{1mm}
\noindent \underline{\textit{Information Seeking communities.}} Human-authored content consistently conveys stronger signals of \textit{knowledge} (i.e., fact-based information) and \textit{similarity} (i.e., drawing parallels between the matter of discussion and personal traits or experiences). This might suggest that humans tend to ground their comments on experiential knowledge, which differs from the ``pre-training'' one achieved by GenAI tools, and with a language that typically aligns with the corresponding community. Conversely, MGT tends to express higher levels of \textit{status} (i.e., expressions of admiration and appreciation) and \textit{support} (i.e., empathy and emotional warmth), likely due to their ``authoritative'' answers and assistive tone that is typical of instruction-tuned chatbots.

\vspace{1mm}
\noindent \underline{\textit{Social Support communities.}} Human-written messages score substantially higher in  \textit{knowledge} and \textit{similarity}. We ascribe the first finding to the hypothesis that supportive knowledge often depends on empathy or experience, which are both unique to humans. Related considerations hold for similarity, where the social and peer support occurring in real life may help in instilling more prominent similarity signals in HGT. On the other hand, MGT tends to contain abundant \textit{support} and \textit{status} expressions, reflecting the optimization (e.g., human preference alignment) of modern GenAI tools in offering instructive and ``comforting'' answers.

\vspace{1mm}
\noindent \underline{\textit{Discussion communities.}} MGT conveys more \textit{knowledge} and \textit{status}. This is likely due to a particularly fertile ground for GenAI, such as producing knowledge-rich content in discussion, and their authoritative styles. HGT, however, exhibits significantly better signals of \textit{support}, \textit{conflict}, and \textit{similarity}. These findings perfectly align with the social dynamics of online debates: users tend to support like-minded peers, engage in confrontational or conflictual interactions, and emphasize shared identities. Notably, the lower conflict scores exhibited by MGT reflect the efforts of preventing such tools from moving towards conflictual behaviors with their users.

\vspace{1mm}
\noindent \underline{\textit{Identity communities.}} MGT is found to convey better signals than HGT only for \textit{status}. This suggests that while GenAI tools might adopt a declarative tone, they struggle in expressing the emotionally and experientially grounded communication typical of identity discourse. Indeed, we found human-authored content to better convey \textit{knowledge}, \textit{fun}, \textit{conflict}, and \textit{similarity}---drawing from personal identity and experience. 

\vspace{1mm}
\noindent \underline{\textit{ChitChat communities.}} We found HGT to outperform MGT in conveying \textit{knowledge} and \textit{similarity}, suggesting a higher capacity to express shared tastes. Again, MGT confirms its ability to convey better \textit{status} and \textit{support}, reflecting its consistent behavior in using explanatory and supportive language even in informal contexts, where these signals are less socially expected.

\subsection{Engagement}
\label{sec:results:engagement}

After characterizing MGT both quantitatively and qualitatively, we examined whether it exhibits systematically different engagement patterns compared to HGT. In particular, we compared the distributions of their \textit{engagement scores} for each \textit{subreddit–month} pair. To ensure meaningful comparisons, we restricted the analysis to cases where both MGT and human comment distributions contained at least 50 samples, resulting in a total of 102 tests.

Following the procedure described in Section~\ref{sec:methods:engagement}, we applied a bootstrap resampling approach ($N=1000$) to balance sample sizes between MGT and HGT comments, performing a Mann–Whitney U test on each resample. For robustness, we considered a difference statistically meaningful only if a majority of bootstrap runs (i.e., more than 50\%) yielded significant results ($p < 0.05$). This filtering step isolates contexts in which divergences between distributions are consistent rather than incidental, yielding 26 significant cases out of 102 evaluated. 

For these statistically significant cases, we complemented the significance testing with an effect size estimation using Cliff’s $\delta$. Remarkably, in all but one case, the effect size was positive ($\delta > 0$), indicating that when significant differences in engagement scores arise, MGT consistently attracts higher engagement than human-authored comments. The only exception was observed in \texttt{r/worldnews} (June 2023), where MGT exhibited significantly lower engagement.

\begin{table}[t]
\centering
\caption{Summary of subreddit–month pairs where engagement distributions between MGT and HGT differ significantly. Positive, resp. negative, values of Cliff’s $\delta$ indicate higher, resp. lower, engagement for MGT. For each case, \%$p$ reports the fraction of significant tests.}
\label{tab:engagement-effects}
\scalebox{0.8}{
\setlength{\tabcolsep}{4pt}
\begin{tabular}{llccc}
\toprule
\textbf{Category} & \textbf{Subreddit} & \textbf{\# Months} & \textbf{Avg. $\delta$} & \textbf{\% $p$} \\
\midrule
\multirow{3}{*}{\textbf{Information Seeking}} 
& \texttt{r/technology} & 3 & 0.27 & 0.72 \\
& \texttt{r/explainlikeimfive} & 1 & 0.17 & 0.55 \\
& \texttt{r/worldnews} & 1 & -0.28 & 0.53 \\
\midrule
\multirow{1}{*}{\textbf{Social Support}} 
& \texttt{r/relationship} & 10 & 0.28 & 0.91 \\
\midrule
\multirow{1}{*}{\textbf{Discussion}} 
& \texttt{r/politics} & 2 & 0.30 & 0.94 \\
\midrule 
\multirow{1}{*}{\textbf{Identity}} 
& \texttt{r/teenagers} & 7 & 0.20 & 0.66 \\
\bottomrule
\end{tabular}
}
\end{table}

Interestingly, the 26 identified subreddit-month pairs in which engagement scores between MGT and HGT differ significantly are concentrated in specific conversational domains (i.e., subreddit categories), as summarized below and in Table~\ref{tab:engagement-effects}.

\begin{itemize}[leftmargin=*]

    \item \textbf{Information-seeking communities}---We found \textbf{five} significant cases in explanatory (\textbf{r/explainlikeimfive}) or technical subreddits (\textbf{r/technology}), with a modest effect size ($\delta \approx 0.20$). Notably, this category yields an interesting outlier, \textbf{r/worldnews} in June 2023, where MGT engagement was significantly lower than for HGT ($\delta = -0.28$), suggesting that users in factual information environments might be more skeptical towards MGT.
   
    \item \textbf{Social-support discussions}---With \textbf{r/relationship\_advice} accounting for \textbf{ten} significant instances between December 2022 and November 2024. These exhibit some of the strongest observed effect sizes ($\delta \approx 0.28$), suggesting that MGT might receive higher engagement in supportive and advice-oriented contexts.

    \item \textbf{Discussion communities}--- \textbf{r/politics} shows \textbf{two} significative instances occurring within August and October 2024, both associated with a relatively large effect ($\delta \approx 0.36$). This suggests that MGT content might attract more engagement during periods of heightened political activity (e.g., 2024 US elections).

    \item \textbf{Identity-oriented spaces}---With \textbf{seven} significant instances (i.e., subreddit-month pairs) primarily in \textbf{r/teenagers} between April 2022 and May 2023, this category accounts for a substantial share of cases. Despite a relatively small effect size ($\delta \approx 0.19$), these are consistently positive, suggesting that MGT tends to receive slightly more engagement than the human counterpart among young people.
\end{itemize}

\section{Discussion}
\label{sec:discussion}

With this first large-scale study of Machine-Generated Text (MGT) on Reddit, we contribute to advancing the understanding of how Generative AI is reshaping discussions and information sharing in online social media. We moved beyond estimating the overall prevalence of MGT and instead examined its \textit{concentration} and \textit{nature} within an online ecosystem. 

Our results show that different communities are affected to varying degrees, with MGT more concentrated in subreddits oriented toward technical knowledge exchange and social support (\textbf{RQ1}). After the public release of ChatGPT, the use of MGT has been steady over time, with peaks of prevalence corresponding to major releases of new LLMs (\textbf{RQ2}). Most of the MGT is concentrated in the activity of a few users, with 2\% of the active users being responsible for the entire MGT production (\textbf{RQ3}). We also found MGT often conveys social signals that differ from those of HGT. Most notably, across many subreddits, MGT exhibits strong patterns of social support and  status-giving, reflecting the characteristic style of modern AI assistants. Finally, MGT comments are often indistinguishable from HGT in terms of positive engagement, and in some cases even outperform them (\textbf{RQ4}).

\vspace{1.5mm}
\noindent 
\textbf{Impact.\ }
These findings have several implications. Practically, they highlight the need for platforms to adapt moderation and transparency policies, particularly in communities where trust and authenticity are central, such as knowledge-sharing and support fora. Theoretically, they suggest that GenAI is not merely amplifying existing discourse but actively shaping new communicative norms, challenging established notions of authenticity and prompting to consider how human and AI voices co-evolve in online ecosystems. Ethically, the stronger engagement elicited by MGT opens questions about the risks of manipulation and representational imbalance.

\vspace{1.5mm}
\noindent 
\textbf{Limitations.\ }
Our contribution has limitations that future work could address. First, we study a single platform, making the findings not generalizable to other social media. Reddit is well-suited for this type of analysis due to its relatively unconstrained comment length, which facilitates MGT detection. Extending the analysis to platforms where textual content is typically much shorter (e.g., $\mathbb{X}$, Bluesky) would require the development of specialized MGT detectors for short-form text---a research challenge in itself.

Second, while our selection of subreddits includes popular, active, and diverse communities, it still represents a small portion of the activity on Reddit. These communities are therefore not necessarily representative of the full spectrum of GenAI use or MGT diffusion across the platform. Nevertheless, by demonstrating that the spread of MGT on Reddit is non-negligible, our study highlights the need for broader, more systematic investigations.

Third, although informative temporal patterns emerge from our analysis, the public adoption of GenAI tools is still too recent to draw any conclusions about long-term trends of MGT use. Within our observation window, MGT prevalence appears relatively stable after the initial surge of adoption between November 2022 and September 2023. However, only a systematic monitoring of the platform over longer periods can shed light on the persistence of this phenomenon. Furthermore, prior work suggests that the diffusion of GenAI may influence communication norms in human-authored text as well~\cite{agarwal2025ai}, raising new questions about the co-evolution of human and AI-generated writing styles that more detailed temporal analyses could help uncover.

Fourth, our comparison of levels of engagement between MGT and HGT is based on a weak form of matching based on a combination of subreddit and month of publication. Stronger forms of matching that account for other factors including topic and the author's popularity would yield a more accurate comparison.

Finally, our adopted detection methodology does not allow us to make claims about the exact prevalence of MGT on Reddit. Our estimates are based on a highly conservative classification threshold ($\tau=0.99$), chosen to demonstrate that even under strict criteria, MGT prevalence is non-negligible in many communities. The actual use of GenAI to produce text on Reddit is likely broader than our estimates suggest. For instance, lowering the classification threshold from $0.99$ to $0.95$ already doubles the estimated prevalence.

\vspace{2mm}
\section{Related Work}
\label{sec:related}

The growing presence of MGT across online platforms has motivated a growing body of research aimed at estimating its prevalence and understanding its potential impact on digital communication. Studies using BERT-based detectors on 15M news articles revealed that by mid-2023, since the release of ChatGPT the proportion of AI-generated news increased by over 57\% on mainstream outlets and by an astonishing 474\% on misinformation and propaganda sites~\cite{hanley2024machine}. Similarly, supervised detectors estimated that AI-generated articles account for roughly 40\% of news contributions on Medium and Quora~\cite{sun2024we}. Analyses of Wikipedia entries based on the GPTZero detector suggest that over 5\% of newly created English articles may be AI-generated~\cite{brooks2024rise}. Even the scientific review process appears affected: recent estimates indicate that between 6\% and 17\% of peer reviews for leading Machine Learning conferences may be produced by LLMs~\cite{liang2024monitoring}.

In contrast, fewer studies have quantified the prevalence of AI-generated text on social media. Earlier research primarily detected machine-generated posts by identifying bot-like authors exhibiting coordinated or automated behaviors~\cite{cinus2025exposing}. Text-based approaches relying on stylometry~\cite{kumarage2023stylometric} offered limited accuracy and generalizability. More recent studies have leveraged crowdsourced signals (such as $\mathbb{X}$'s Community Notes) to examine the diffusion and nature of AI-generated discourse~\cite{drolsbach2025characterizing}.

On Reddit, community responses to AI-generated content have been characterized by skepticism and concern~\cite{lloyd2023there}, leading moderators and administrators to introduce new governance rules soon after the public release of ChatGPT~\cite{lloyd2025ai}. Crowdsourced data indicate that AI-generated visual content remained relatively rare through late 2023, representing fewer than 0.5\% of image-based posts~\cite{matatov2024examining}. In this context, the most closely related study to ours is the recent work by Sun et al.~\cite{sun2024we}, who applied a supervised detection method to 982K Reddit posts published between January 2022 and July 2024, estimating that around 2.5\% were AI-generated. However, their analysis focuses solely on prevalence and does not investigate the nature, context, or engagement dynamics of machine-generated content across communities---gaps that our work seeks to address.

\vspace{1mm}
\section*{Acknowledgements}
AT, resp. LLC, is supported by project  ``Future Artificial Intelligence Research (FAIR)'' spoke 9  (H23C22000860006), resp.   project SERICS (PE00000014), both  under the MUR National Recovery and Resilience Plan funded by  the EU - NextGenerationEU. 
LMA acknowledges the support from the Carlsberg Foundation through the COCOONS project (CF21-0432).

\clearpage

\bibliographystyle{ACM-Reference-Format}
\bibliography{sample-base}

\clearpage

\appendix

\setcounter{figure}{0}
\setcounter{table}{0}
\setcounter{equation}{0}
\renewcommand{\thefigure}{A\arabic{figure}}
\renewcommand{\thetable}{A\arabic{table}}
\renewcommand{\theequation}{A\arabic{equation}}

\section{Additional Details on Methodology}

\subsection{Categorization of Reddit Communities}
\label{app:taxonomy}
Below, we provide the full list of our 51 selected subreddits across the five community types:

{\raggedright
\vspace{1.5mm}
\noindent \textbf{Information Seeking.\ }
r/worldnews, r/DIY, r/askscience, r/personalfinance, r/technology, r/history, r/CryptoCurrency, r/nutrition, r/learnprogramming, r/explainlikeimfive, r/health, r/techsupport, r/writing.

\vspace{1.5mm}
\noindent \textbf{Social Support.\ }
r/LifeProTips, r/GetMotivated, r/relationship\_advice, r/lifehacks, r/malefashionadvice, r/dating, r/femalefashionadvice, r/careerguidance, r/loseit, t/tifu, r/Parenting, r/fitness, r/AITAH, r/AmIOverreacting, r/offmychest, r/SuicideWatch, r/vent, r/mentalhealth.

\vspace{1.5mm}
\noindent \textbf{Discussion.\ }
r/politics, r/unpopularopinion, r/changemyview, r/SeriousConversation, r/PoliticalDiscussion.

\vspace{1.5mm}
\noindent \textbf{Identity.\ }
r/BlackPeopleTwitter, r/AskWomen, r/IAmA, r/teenagers, r/introvert, r/asktransgender, r/Conservative, r/Liberal, r/Libertarian, r/Socialism.

\vspace{1.5mm}
\noindent \textbf{ChitChat.\ }
r/funny, r/Showerthoughts, r/entertainment, r/books, r/CasualConversation.
}

\subsection{Details on Experimental Setup}
\label{app:exp-setup}
All experiments have been performed on an 8x NVIDIA A30 GPU 24GB NVRAM server with 24 GB of RAM each, 764 GB of system RAM, a Double Intel Xeon Gold 6248R with a total of 96 cores, and  Ubuntu Linux 20.04.6 LTS as OS.

To ease reproducibility, we adopted Fast-DetectGPT via its default parameter settings, as in their official GitHub repository.\footnote{\url{https://github.com/baoguangsheng/fast-detect-gpt}}

\subsection{Temporal Distribution of Analyzed Data}
\label{app:temporal-distribution-comments}

\begin{figure}[t!]
    \centering
    \includegraphics[width=0.9\linewidth]{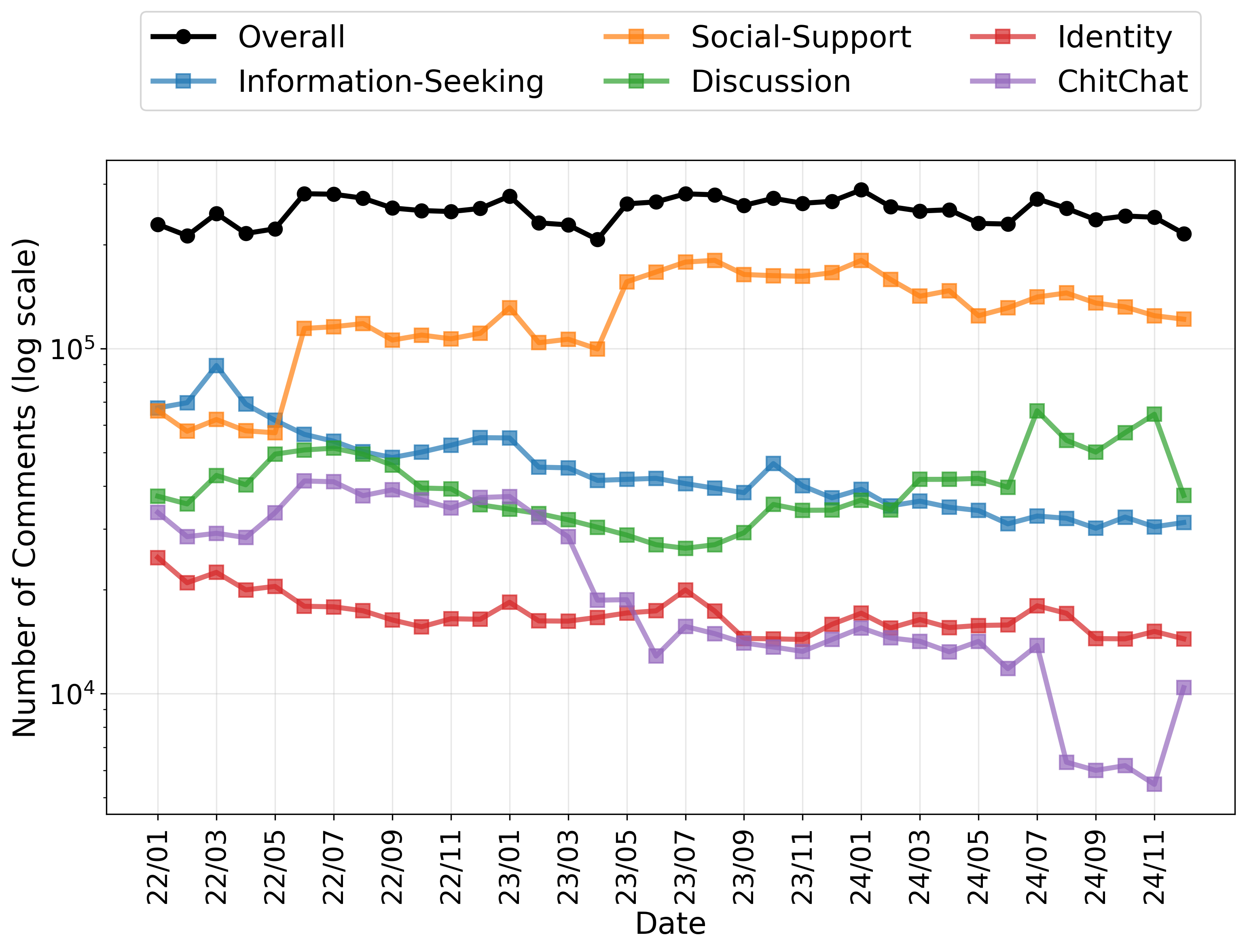}
    \caption{Monthly volume of comments for each subreddit category.}
    \label{fig:appendix-temporal-distrib}
\end{figure}

\begin{figure}[t!]
    \centering
    \includegraphics[width=0.9\linewidth]{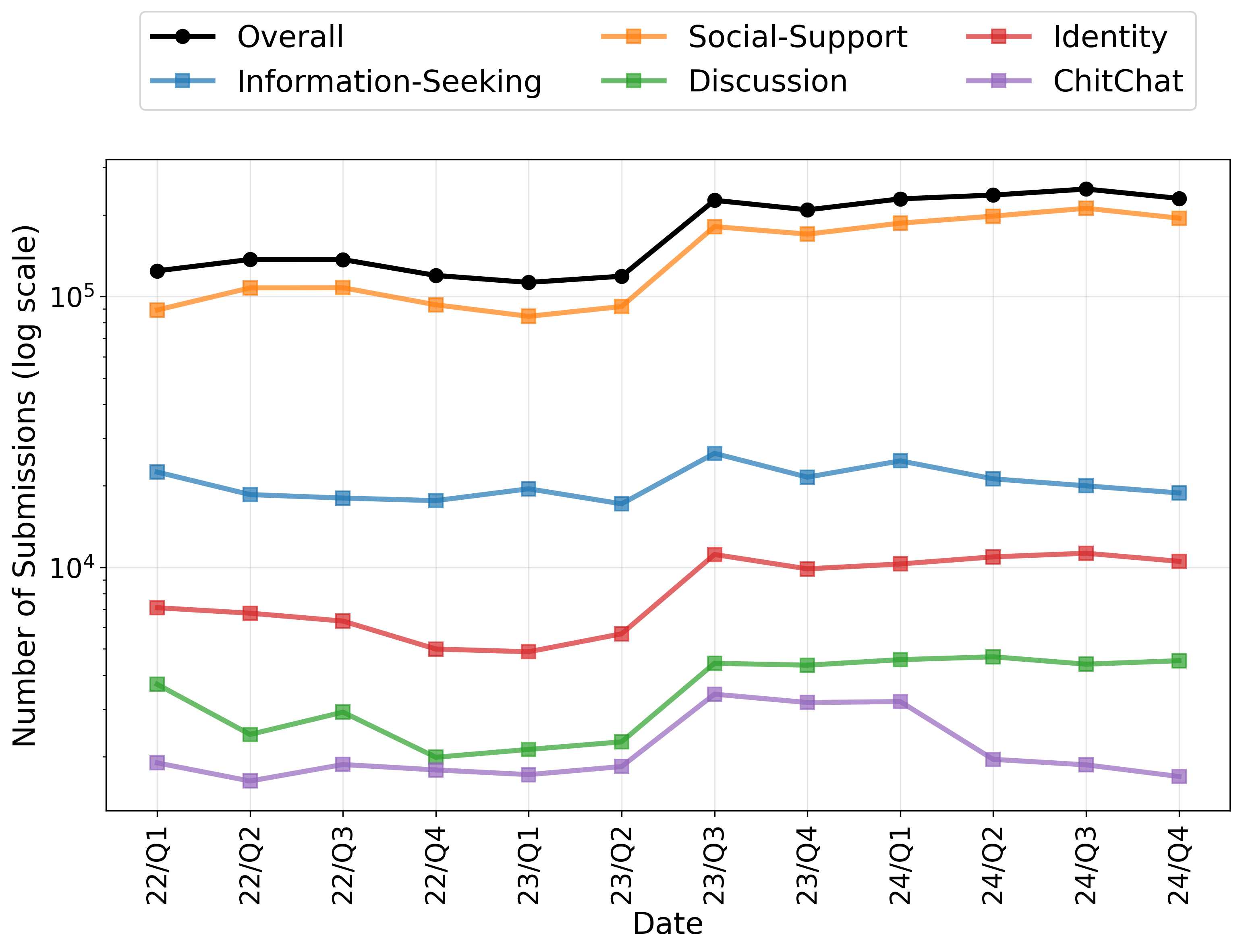}
    \caption{Quarterly volume of submissions for each subreddit category.}
    \label{fig:appendix-temporal-distrib-submissions}
\end{figure}

Figures~\ref{fig:appendix-temporal-distrib}-\ref{fig:appendix-temporal-distrib-submissions} report the number of comments and submissions over time, across the different community types. Submission data is aggregated by quarters instead of monthly to counter sparsity.

\subsection{Details on Social Dimensions}
\label{app:socialdimensions}

Table~\ref{tab:socialdims-description} reports the short description of the social dimensions that we quantify in our study, as defined by Choi et al.~\cite{Choi20tensocial}.

\begin{table}
    \centering
    \caption{Summary characteristics of Social Dimensions, as originally reported in \cite{Choi20tensocial}.}
    \label{tab:socialdims-description}
    \begin{tabular}{lp{0.7\columnwidth}}
    \toprule
        \textbf{Dimension} & \textbf{Description} \\
    \midrule
        \textbf{Knowledge} & Exchange of ideas or information; learning, teaching \\
        \textbf{Status} & Conferring status, appreciation, gratitude, or admiration upon another \\
        \textbf{Support} & Giving emotional or practical aid and companionship \\
        \textbf{Fun} & Experiencing leisure, laughter, and joy \\
        \textbf{Conflict} & Contrast or diverging views \\
        \textbf{Similarity} & Shared interests, motivations or outlooks\\
    \bottomrule
    \end{tabular}
\end{table}

\clearpage

\section{Machine-Generated Text in Submissions}
\label{app:submissions-results}

\begin{figure}[t!]
    \centering
    \includegraphics[width=\linewidth]{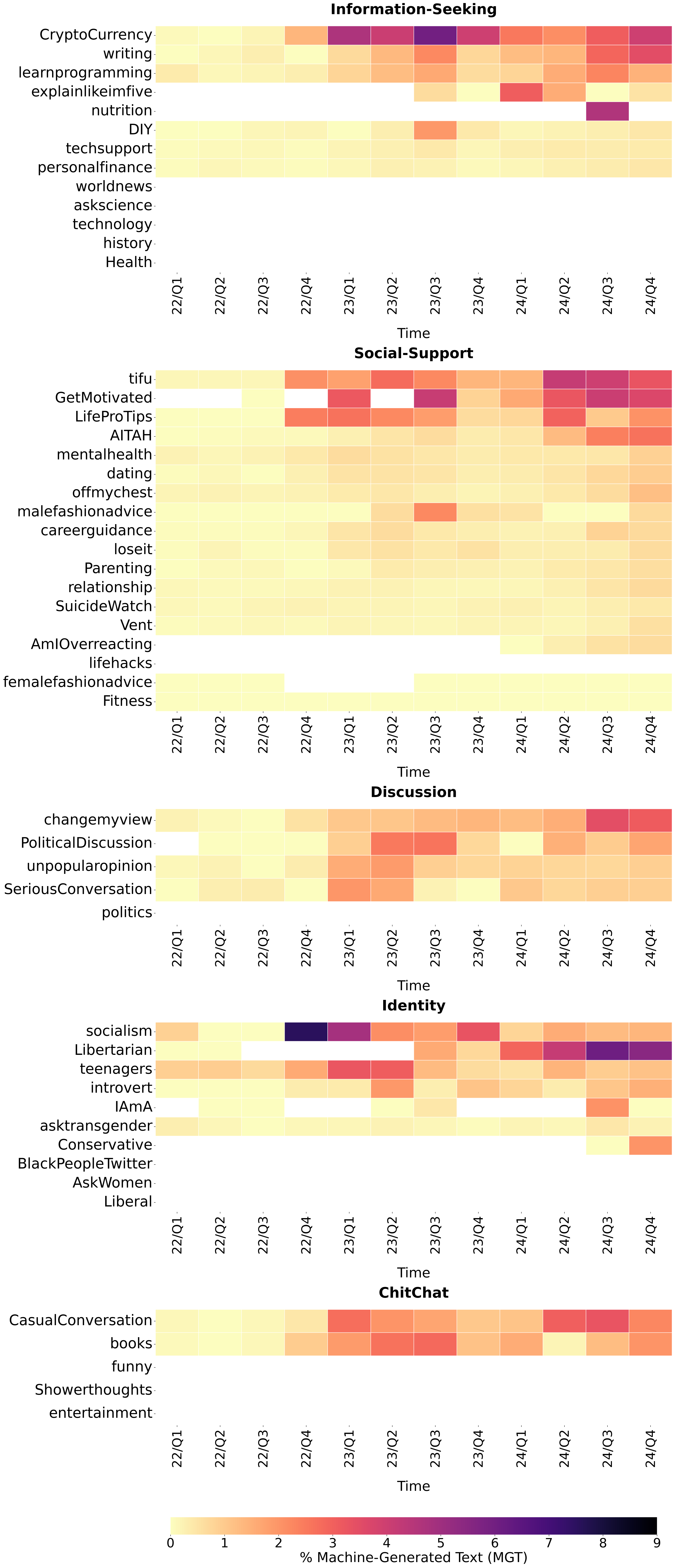}    
    \caption{MGT submission usage across subreddit categories. Rows are sorted by total quarterly adoption. Each cell shows the share of MGT posts for a subreddit-quarter pair. Darker shades indicate higher usage. Only subreddits with $\geq 1$ post/day on avg. are shown; empty cells mean no matching submissions.}
    \label{fig:rq1-submissions}
\end{figure}

\begin{figure*}
    \centering
    \includegraphics[width=1\linewidth]{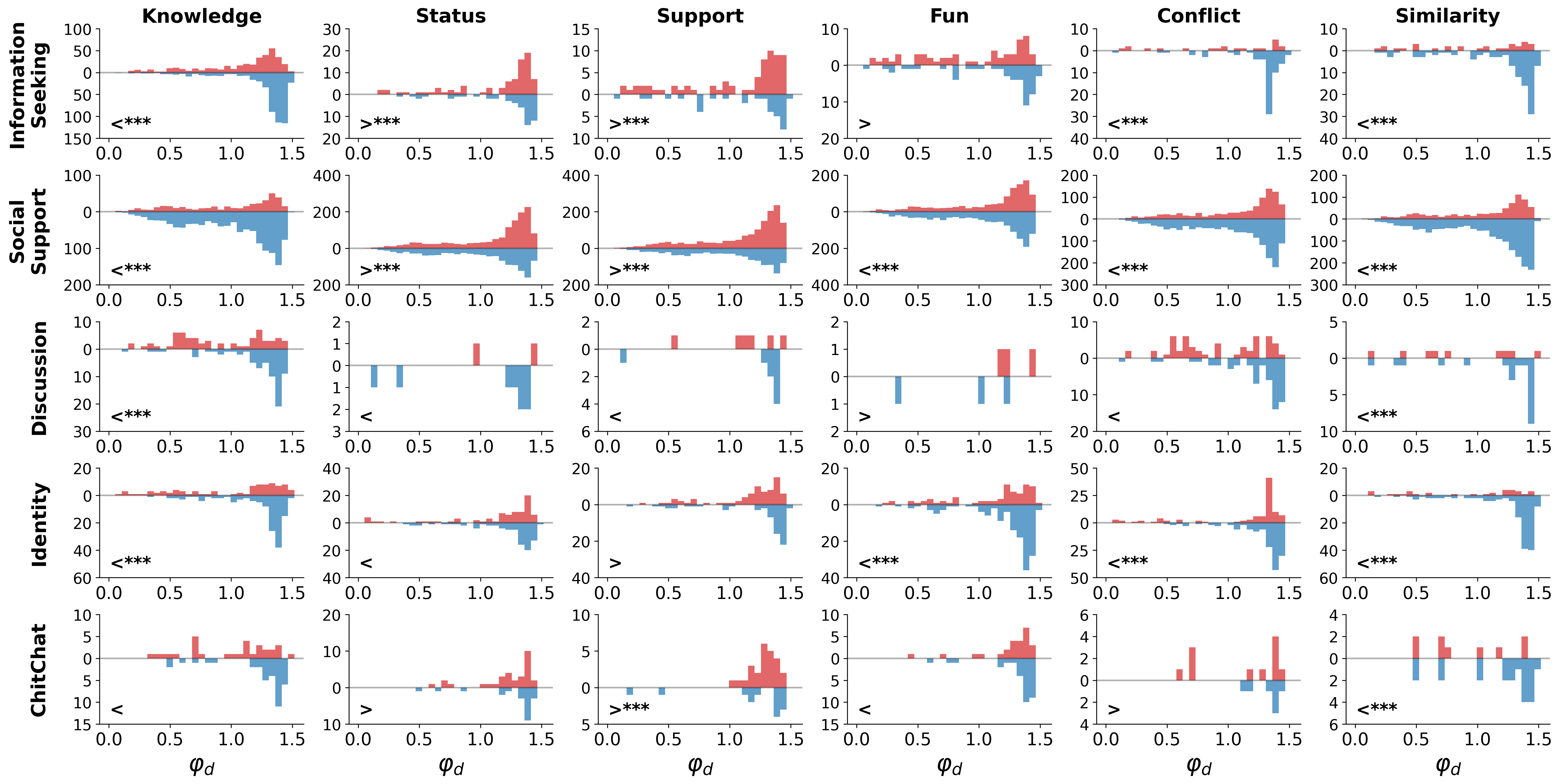}
    \caption{Comparison of social dimension expressions between MGT (in red) and HGT (in blue) submissions across different subreddit categories. Each subplot is labeled with the corresponding effect size, where $>$ indicates that MGT scores higher than HGT, and $<$ indicates the opposite. Marker $^{***}$   denotes that the difference in social dimension expression scores between MGT and HGT is statistically significant according to the Mann–Whitney U test. }
    \label{fig:tendims-submissions}
\end{figure*}

We applied to Reddit submissions the same analytical framework used for comments. Figure~\ref{fig:rq1-submissions} illustrates the prevalence of MGT across subreddits and over time, while Table~\ref{tab:mgt-peaks} summarizes statistics on peak prevalence across subreddit types. Table~\ref{tab:statistics-submissions} reports measures of text length, readability, and compressibility for submissions, and Figure~\ref{fig:tendims-submissions} shows the distribution of social dimensions for MGT and HGT. Engagement statistics are presented in Table~\ref{tab:engagement-effects-submissions}. All measurements are aggregated by quarters instead of monthly to counter sparsity.

Overall, the patterns observed in submissions closely resemble those found in comments. The estimated prevalence of MGT remains stable over time since late 2022, with peaks reaching approximately 7\%, and is similarly concentrated among a small number of authors. As with comments, machine-generated submissions tend to be longer and less readable than their human-authored counterparts. MGT also conveys more signals of social support but fewer of knowledge exchange, conflict, or similarity.

We identify three main differences compared to comments. First, the subreddits with the highest prevalence of MGT differ slightly: r/CryptoCurrency, r/tifu, r/changemyview, and r/socialism show higher MGT prevalence in submissions but only modest levels in comments. Notably, all Discussion communities exhibit more MGT in submissions than in their corresponding comment sections. Second, MGT submissions tend to convey less status-related intent than comments (particularly in Discussion communities), likely because submissions lack the conversational dynamics that might naturally induce expressions of appreciation. Third, although engagement levels for MGT submissions are generally indistinguishable from those of HGT (as observed for comments), when significant differences do occur, they tend to disfavor MGT. This may reflect the greater visibility and scrutiny of submissions compared to comments, which could penalize text perceived as less authentic or of lower quality.

\begin{table}[t]
\centering
\caption{Peak prevalence of MGT across subreddit categories. For each category, we report the fraction of MGT submissions over the total detected, the maximum observed share of MGT, the subreddit in which it occurred, and the corresponding date (yy/quarter).}
\label{tab:mgt-peaks-submissions}
\scalebox{0.9}{
\begin{tabular}{lrrlc}
\toprule
\textbf{Community} & \textbf{MGT \%} & \textbf{Peak \%} & \textbf{Top Subreddit} & \textbf{Date} \\
\midrule
Inf. Seek.   & 14.23      & 6.00 & \texttt{r/CryptoCurrency}          & 23/Q3 \\
Soc. Supp.   & 71.42       & 4.16 & \texttt{r/tifu}   & 24/Q2 \\
Discussion   & 3.72      & 3.45 & \texttt{r/changemyview}            & 24/Q3 \\
Identity     & 7.26       & 7.53 & \texttt{r/socialism}           & 22/Q4 \\
ChitChat     & 3.37       & 3.30 & \texttt{r/CasualConversation}               & 24/Q3 \\
\bottomrule
\end{tabular}
}
\end{table}

\begin{figure}[t!]
    \centering
    \includegraphics[width=1\linewidth]{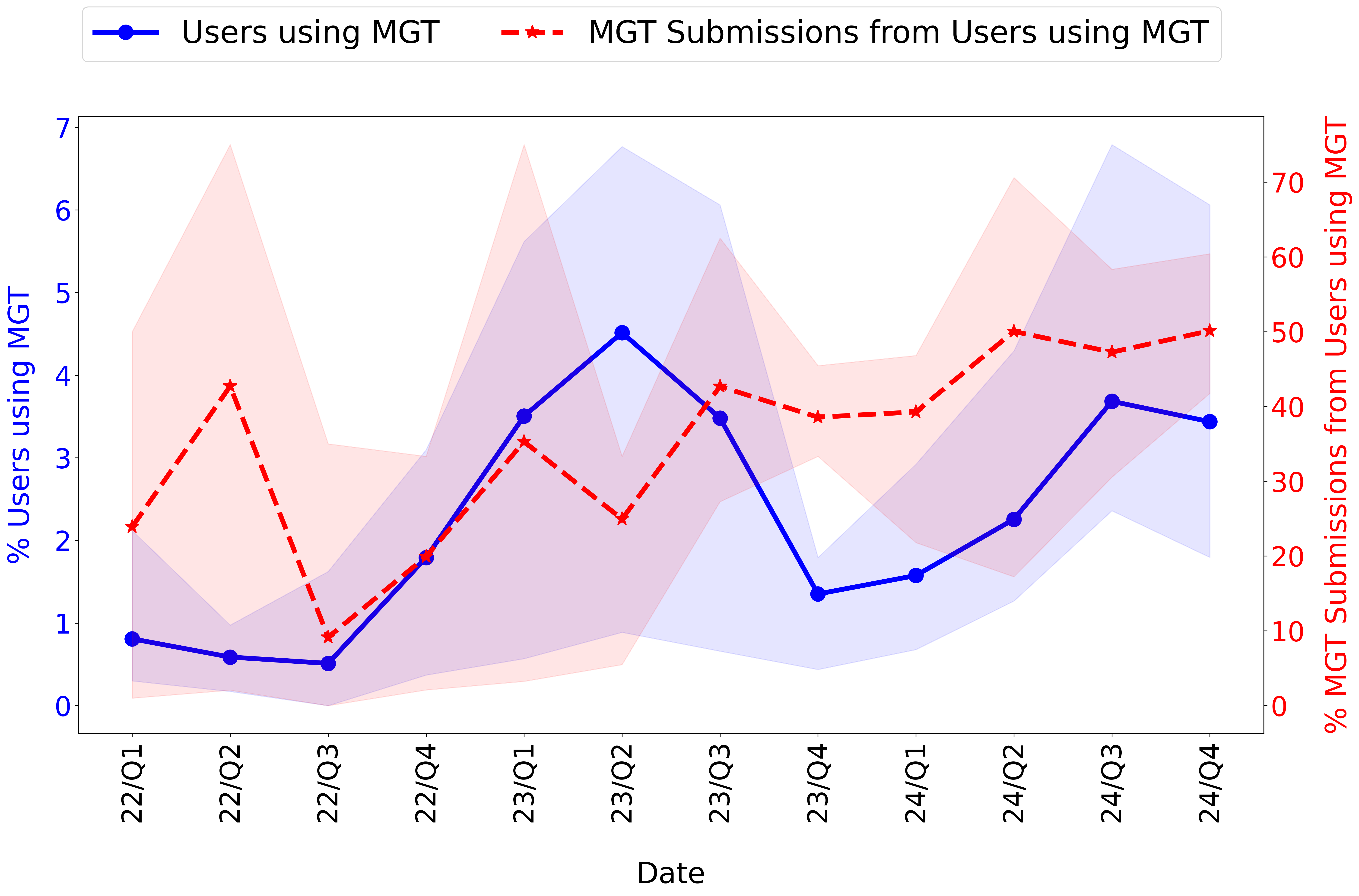}
    \caption{Average number of users adopting MGT (in blue, left y-axis) and the corresponding percentage of their submissions detected to be MGT (in red, right y-axis) across all categories over time. Shaded areas denote minimum and maximum observed values. Users with a single post have been filtered out to mitigate noise due to one-off activities.}
    \label{fig:users_posts_aiusage-submissions}
\end{figure}

\begin{table}[t!]
\caption{Aggregated values of text statistics from the machine-generated and human-authored submissions across subreddit categories. All category–statistics combinations (but \# Words and Compression for Social Support) exhibit statistically significant distributional differences ($p< 0.05$) according to the Mann–Whitney U test.}
\label{tab:statistics-submissions}
\setlength{\tabcolsep}{1.8pt}
\scalebox{0.8}{
\begin{tabular}{lc|r|r|r|r}
\toprule
\textbf{Category} & \textbf{MGT} 
& \multicolumn{1}{c|}{\textbf{\# Words}} & \multicolumn{1}{c|}{\textbf{\# Sentences}} & \multicolumn{1}{c|}{\textbf{Flesch Read.}} & \multicolumn{1}{c}{\textbf{Compression}} \\
\midrule
\multirow{2}{*}{\textbf{Inf.Seek.}} & \xmark & 352.8 ± 338.95 & 20.79 ± 22.12 & 58.3 ± 28.88 & 3.07 ± 0.72 \\
& \cmark & 527.16 ± 488.73 & 31.49 ± 34.13 & 51.59 ± 21.63 & 3.48 ± 0.57 \\
\midrule

\multirow{2}{*}{\textbf{Soc.Supp.}} & \xmark & 493.05 ± 357.3 & 24.78 ± 18.87 & 71.96 ± 51.61 & 3.17 ± 0.42 \\
& \cmark & 471.2 ± 337.14 & 28.52 ± 20.82 & 74.86 ± 18.06 & 3.2 ± 1.27 \\
\midrule

\multirow{2}{*}{\textbf{Discussion}} & \xmark & 404.86 ± 400.81 & 20.31 ± 22.08 & 56.92 ± 26.25 & 3.11 ± 0.31 \\
& \cmark & 650.27 ± 443.12 & 34.45 ± 22.75 & 43.06 ± 28.13 & 3.51 ± 0.42 \\
\midrule

\multirow{2}{*}{\textbf{Identity}} & \xmark & 356.66 ± 234.28 & 17.39 ± 15.16 & 59.95 ± 80.08 & 3.53 ± 8.47 \\
& \cmark & 552.99 ± 802.74 & 33.97 ± 86.57 & -10.09 ± 484.72 & 10.03 ± 48.86 \\
\midrule

\multirow{2}{*}{\textbf{ChitChat}} & \xmark & 345.11 ± 266.0 & 20.08 ± 19.77 & 74.36 ± 12.13 & 2.96 ± 0.29 \\
& \cmark & 430.22 ± 381.33 & 23.84 ± 18.42 & 61.45 ± 29.28 & 3.19 ± 0.46 \\

\bottomrule

\end{tabular}
} 
\end{table}

\begin{table}[t]
\centering
\caption{Summary of subreddit–quarter pairs where engagement distributions between MGT and HGT differ significantly. Positive, resp. negative, values of Cliff’s $\delta$ indicate higher, resp. lower, engagement for MGT. For each case, \%$p$ reports the fraction of significant tests.}
\label{tab:engagement-effects-submissions}
\scalebox{0.8}{
\setlength{\tabcolsep}{4pt}
\begin{tabular}{llccc}
\toprule
\textbf{Category} & \textbf{Subreddit} & \textbf{\# Quarters} & \textbf{Avg. $\delta$} & \textbf{\% $p$} \\
\midrule
\multirow{2}{*}{\textbf{Information Seeking}} 
& \texttt{r/CryptoCurrency} & 4 & -0.29 & 0.97 \\
& \texttt{r/writing} & 2 & -0.25 & 0.86 \\
\midrule
\multirow{3}{*}{\textbf{Social Support}} 
& \texttt{r/AITAH} & 4 & 0.15 & 0.88 \\
& \texttt{r/offmychest} & 2 & 0.24 & 0.61 \\
& \texttt{r/tifu} & 1 & -0.26 & 0.89 \\
\midrule
\multirow{1}{*}{\textbf{Discussion}} 
& \texttt{r/changemyview} & 1 & -0.21 & 0.65 \\
\bottomrule
\end{tabular}
}
\end{table}

\end{document}